# Enhanced coherence of all-nitride superconducting qubits epitaxially grown on silicon substrate

Sunmi Kim[1,*], Hirotaka Terai[2,†], Taro Yamashita[3], Wei Qiu[2], Tomoko Fuse[1], Fumiki Yoshihara[1], Sahel Ashhab[1], Kunihiro Inomata[4], and Kouichi Semba[1,5,‡]

[1]Advanced ICT Research Institute, National Institute of Information and Communications Technology (NICT), 4-2-1, Nukui-Kitamachi, Koganei, Tokyo 184-8795, Japan
[2]Advanced ICT Research Institute, NICT, 588-2 Iwaoka, Nishi-ku, Kobe, Hyogo 651-2492, Japan
[3]Graduate School of Engineering, Nagoya University, Furo-cho, Chikusa-ku, Nagoya, Aichi 464-8603, Japan
[4]Research Center for Emerging Computing Technologies, Advanced Industrial Science and Technology (AIST), Central 2, 1-1-1 Umezono, Tsukuba, Ibaraki 305-8568, Japan
[5]present address: Institute for Photon Science and Technology, The University of Tokyo, Tokyo 113-0033, Japan

[*]kimsunmi@nict.go.jp, [†]terai@nict.go.jp, [‡]semba@ipst.s.u-tokyo.ac.jp

## Abstract

Improving the coherence of superconducting qubits is a fundamental step towards the realization of fault-tolerant quantum computation. However, coherence times of quantum circuits made from conventional aluminium-based Josephson junctions are limited by the presence of microscopic two-level systems in the amorphous aluminum oxide tunnel barriers. Here, we have developed superconducting qubits based on NbN/AlN/NbN epitaxial Josephson junctions on silicon substrates which promise to overcome the drawbacks of qubits based on Al/AlOx/Al junctions. The all-nitride qubits have great advantages such as chemical stability against oxidation, resulting in fewer two-level fluctuators, feasibility for epitaxial tunnel barriers that reduce energy relaxation and dephasing, and a larger superconducting gap of ~5.2 meV for NbN, compared to ~0.3 meV for aluminium, which suppresses the excitation of quasiparticles. By replacing conventional MgO by a silicon substrate with a TiN buffer layer for epitaxial growth of nitride junctions, we demonstrate a qubit energy relaxation time $T_1 = 16.3~\mu s$ and a spin-echo dephasing time $T_2 = 21.5~\mu s$. These significant improvements in quantum coherence are explained by the reduced dielectric loss compared to the previously reported $T_1 \approx T_2 \approx 0.5~\mu s$ of NbN-based qubits on MgO substrates. These results are an important step towards constructing a new platform for superconducting quantum hardware.

**Introduction**

Since the first demonstration of nanosecond-scale quantum coherence in a charge qubit in 1999[1], superconducting qubits have developed into a leading platform for scalable quantum computing, as evidenced by recent key demonstrations including quantum algorithms[2,3], quantum error correction[4-6] and quantum supremacy[7]. However, their gate fidelities still need to be improved further for building a fault-tolerant quantum computer, even though two-qubit gates with fidelities as high as 99.7%[8] have recently been achieved (cf. the case of trapped-ion qubits above 99.9%[9]).

The remarkable progress of superconducting qubits has to a large extent resulted from the innovative five-order increase in their coherence times by improving circuit designs, materials and fabrication processes[10-12]. Looking at material-based improvements, most of the research has been focused on materials for capacitors or microwave resonators (e.g., niobium titanium nitride (NbTiN)[13], TiN[14], and tantalum (Ta)[15] instead of the commonly used niobium (Nb) or aluminum (Al)) in order to reduce microwave dielectric loss induced by uncontrolled defects in oxides at their surfaces and interfaces. In contrast, alternative materials for the Josephson junctions (JJs) of the qubits have not been studied adequately, even though it has been pointed out that the coherence times of superconducting quantum circuits made from conventional Al-based JJs are limited by energy or phase relaxation due to microscopic two-level systems (TLSs) in the amorphous aluminum oxide ($AlO_x$) tunnel barriers[16-18]. Therefore, a breakthrough in the further improvement of superconducting qubits can be expected by exploring alternative materials for JJs.

One promising approach to reduce the number of TLSs in a JJ is to use a crystalline tunnel barrier as demonstrated with epitaxial $Al_2O_3$ layers grown on crystalline Re films[19,20]. From this viewpoint, fully epitaxial nitride JJs consisting of NbN/AlN/NbN tri-layers are highly attractive candidates as an alternative material technology for superconducting quantum circuits. This is because such epitaxially grown nitride JJs have great potential to solve material-related concerns, including the TLS problem, owing to their high crystal quality and chemical stability against oxidation. Moreover, due to the large superconducting gap ($2\Delta \sim 5.2$ meV) and relatively high transition temperature ($\sim 16$ K) of NbN[21], quasiparticle excitation can be suppressed. One point that requires attention when using an AlN tunnel barrier is its piezoelectric property, i.e. the coupling between the electric field concentrated at the JJ and the underlying crystal lattice, which leads to substantial qubit decoherence via phonon emission[22]. In previous studies of phase qubits using an amorphous AlN barrier, energy relaxation times were as short as ~10 ns due to piezoelectricity[23,24]. However, the detrimental piezoelectric effect can be avoided when the AlN barrier is grown epitaxially in the cubic phase because of its lattice-inversion symmetry. In an early attempt with epitaxially grown cubic-phase AlN tunnel barriers[25], longer relaxation times of 500 ns were observed in transmon qubits consisting of fully epitaxial NbN/AlN/NbN JJs on a single-crystal MgO substrate, commonly used because it allows good lattice matching between the materials. These coherence times were limited by the dielectric loss from the MgO substrate rather than the intrinsic properties of the NbN-based JJs themselves. In the context of NbN-based qubits, it is worth mentioning that an early report of a phase qubit using an epitaxially grown 10 μm-size NbN/AlN/NbN JJ also showed a relatively long phase coherence time of about 5 μs[26]. It has been characterized using a direct current (DC) readout so that the phase qubit is quite insensitive to the dielectric loss of substrate, which explains the relatively long coherence times obtained in that work.

Yet, the development of NbN-based qubits is still inadequate despite the superior properties of NbN as an alternative material for superconducting quantum hardware. More research is required in this direction to go beyond conventional Al-based qubits.

In this article, we report a significant improvement in quantum coherence of an epitaxially grown NbN-based superconducting qubit. To suppress the dominant dielectric loss from the MgO substrate, which has limited the energy relaxation time ($T_1$) in the previous work[25], we adopted a Si substrate with a TiN buffer layer for epitaxial growth of NbN/AlN/NbN junctions[27]. The loss tangent of the high-resistivity Si substrate is much lower than that of MgO, as reported in Ref. 28 ($5 \times 10^{-6} - 12 \times 10^{-6}$ for the former and $0.5 \times 10^{-4} - 0.8 \times 10^{-4}$ for the latter). Additionally, for the qubit design, we employed the structure of a capacitively-shunted (C-shunt) flux qubit, which brings several advantages such as enhanced coherence and reproducibility as well as higher anharmonicity compared to transmon qubits[29-31]. With these modifications, we achieved coherence times as long as tens of microseconds using an all-nitride qubit with epitaxial JJs on a Si substrate. These coherence times are more than an order of magnitude longer than those obtained with qubits grown on MgO substrates. We explain that this improvement in coherence times can be attributed mainly to the reduced dielectric loss when replacing the MgO substrate with the Si substrate. These results shed light on the role of alternative material technologies in improving superconducting quantum circuits

## Results and discussions

**All-nitride C-shunt flux qubit.** Our circuit is made of epitaxially grown NbN with a TiN buffer layer on a Si substrate. It consists of a C-shunt flux qubit that is based on epitaxial NbN/AlN/NbN JJs and is capacitively coupled to a half-wavelength superconducting coplanar waveguide resonator (CPW) (see Fig. 1). A detailed experimental setup together with device parameters are found in the Methods.

**Spectroscopy of resonator.** First, we assess the resonator properties as shown in Fig. 2(a). The resonator's transmission parameter ($S_{21}$) is plotted as a function of the probe frequency and the normalized flux bias $\Phi/\Phi_0$, where $\Phi$ is the magnetic flux bias applied through the qubit loop and $\Phi_0$ is the superconducting flux quantum. The power applied to the resonator was -138 dBm, corresponding to the level of a single photon in the resonator. The resonator has a fundamental resonance frequency $\omega_r/2\pi = 9.796$ GHz at the qubit flux bias point $\Phi/\Phi_0 = 0.5$ and a loaded quality factor $Q_L = 1.405 \times 10^4$ extracted from the full width at half maximum of the power spectrum fitted by a Lorentzian function with linewidth $\kappa/2\pi = 0.697$ MHz (Fig. 2(b)). An internal quality factor $Q_{int} = 2.72 \times 10^5$ can be obtained from the relation[32] $Q_{int} = Q_L/(1 - 10^{-IL/20})$, where IL=0.461 dB is the insertion loss measured under careful calibration. The corresponding loss tangent of our NbN resonator fabricated on a Si substrate is $\tan\delta = 1/Q_{int} = 3.68 \times 10^{-6}$, which is almost two orders of magnitude lower than that of the MgO substrate (i.e. $\tan\delta_{(MgO)} = 3.30 \times 10^{-4}$)[33]. On both sides of $\Phi/\Phi_0 = 0.5$ flux bias in Fig. 2(a), clear anti-crossings are observed at the points where the qubit energy is equal to the resonator's photon energy, and the vacuum Rabi splitting is found to be ~120 MHz.

**Spectroscopy of qubit.** The measured transition frequency between the qubit's ground and first excited state ($\omega_{01}/2\pi$) is shown in Fig. 2(c). At the flux-insensitive point (i.e., $\Phi/\Phi_0 = 0.5$), the qubit has its minimum frequency, 6.61 GHz, which is detuned by $(\omega_r - \omega_{01})/2\pi = 3.19$ GHz from the fundamental resonator mode. As shown by the dashed line in Fig. 2(c), the qubit spectrum $\omega_{01}/2\pi$ is reproduced by simulations with fitting parameters such as the area ratio of the small JJ relative to the two larger JJs $\alpha = 0.358$, Josephson energy $E_J/h = 140$ GHz (where $h$ is the Plank constant, and the corresponding critical current density of the larger junctions in the qubit is 38.6 A cm$^{-2}$), and charging energy $E_C/h = (e^2/2C_\Sigma)/h = 0.244$ GHz. Here, the total qubit capacitance ($C_\Sigma$) is 79.6 fF which includes a shunt capacitance $C_S = 52.8$ fF and the total junction capacitance of the flux qubit $C_J = 26.8$ fF (the detailed parameters are also found in the Methods).

**Energy relaxation time, $T_1$, and its temporal variation.** Qubit coherence properties were obtained from time-domain measurements. At the flux-insensitive point, we measured the energy relaxation time ($T_1$), Ramsey decay time ($T_2^*$), and spin-echo coherence time ($T_2$) by applying the corresponding control-pulse sequences (see the insets in Figs. 3(a), 4(a) and 4(b)). The qubit's excited state population is measured by a digitizer and ensemble-averaged over $6.5 \times 10^4$ repetitions. The resulting signal is plotted as a function of the delay time ($\tau$) in Figs. 3 and 4. Figure 3(a) shows the data for energy relaxation, which is well fitted by an exponential decay function $\exp(-\tau/T_1)$, giving $T_1 = 18.25 \pm 0.91$ μs. In addition, we repeated the $T_1$ measurement 100 times over a period of 33 hours to observe the $T_1$ fluctuations of our nitride qubit (see Fig. 3(b)). The obtained values range from 8 to 20 μs, and the histogram of the $T_1$ data is well fitted by a Gaussian distribution with peak center $\bar{T}_1 = 16.3$ μs and standard deviation $\sigma_{T1} = 1.73$ μs as shown in Fig. 3(c). These value of $T_1$ are the highest among all-nitride qubits. Currently they are lower than those of Al-based C-shunt flux qubits coupled to two-dimensional resonators on Si substrates[30] (showing $T_1$=55 μs). However, there is still room for improvements in a number of aspects to increase $T_1$ further, and the significant improvement over state-of-the-art all-nitride qubits is an important step in that direction.

The temporal variation of $T_1$ is usually explained by quasiparticle fluctuations[30] and instability of TLS defects[34,35]. When compared to Al-based single-junction Xmon-type transmon qubits in Ref. 35 (where $T_1$ histograms show rather broad Gaussian distributions with $\sigma_{T1} \sim 20\%$ of $\bar{T}_1$, with parameters $\bar{T}_1 = 46.18$ μs and $\sigma_{T1} = 10.24$ μs for one sample, and $\bar{T}_1 = 70.72$ μs and $\sigma_{T1} = 14.31$ μs for the other sample), we found that the $T_1$ data of our nitride qubit show little temporal variation, $\sigma_{T1} \sim 10\%$ of $\bar{T}_1$. Such a narrow Gaussian distribution is also observed in Al-based C-shunt flux qubits[30] and discussed as an indication that quasiparticles did not strongly influence this device. We therefore believe that our nitride qubit is also not strongly affected by quasiparticles. The instances of large deviation in $T_1$ to lower values outside the Gaussian peak in Fig. 3(c), i.e., the outliers, can be explained by weakly coupled TLS defects in the remaining silicon dioxide (SiO$_2$) after buffered hydrogen fluoride (BHF) treatment in our fabrication process. To reach a quantitative understanding of the degree to which two-level system could be limiting the qubit coherence time, more experiments are needed. Such an investigation is beyond the scope of this work and could be the subject of future work.

**Phase relaxation times, $T_2^*$ and $T_2$.** We have measured the coherence times for phase relaxation from Ramsey and spin-echo experiments as shown in Fig. 4. The signals, which oscillate at the detuning frequency $\delta\omega = |\omega_{01} - \omega_d|/2\pi = 5$ MHz (where $\omega_d/2\pi$ is the drive frequency), can be fitted by exponentially decaying sinusoidal functions with relaxation times $T_2^* = 3.33 \pm 0.30$ μs and $T_2 = 23.2 \pm 5.21$ μs (see Fig 4(a) and (b)). The $T_2^*$ and $T_2$ measurements are also repeated 100 times along with the $T_1$ measurements. The resulting histograms are shown in Figs. 4(c) and (d). The observed values of $T_2^*$ are in the range $1.2 - 4.4$ μs, and the corresponding Gaussian fit has a center value $\bar{T}_2^* = 3.25$ μs and a standard deviation $\sigma_{T2*} = 0.44$ μs. The $T_2$ values, obtained by applying an additional $\pi$ pulse between the $\pi/2$ pulses to decouple low frequency noise, are remarkably higher and lie in the range $14 - 41$ μs. The obtained center value of the Gaussian distribution is $\bar{T}_2 = 21.5$ μs.

**Main factors behind the enhanced coherence time.** Compared with the NbN-based qubit epitaxially grown on a MgO substrate ($T_1 \approx T_2 \approx 0.5$ μs)[25], the obtained coherence times for the qubit on a Si substrate ($\bar{T}_1 = 16.3$ μs and $\bar{T}_2 = 21.5$ μs as the mean values) represent a significant improvements, namely 32-fold increase in $T_1$ and a 43-fold increase in $T_2$. To support the argument that the substrate material is responsible for this improvement, we briefly discuss the predicted limit on the quality factor for the MgO substrate material. As reported in Ref. 25, the predicted limit for the MgO substrate clearly appears in the internal quality factor of the resonator $Q_{int} \sim 2.35 \times 10^3$. This value of $Q_{int}$ gives a higher value of the loss tangent for the MgO substrate $\tan\delta \sim 1/Q_{int} \sim 4.25 \times 10^{-4}$ compared to our result for the Si substrate, i.e., $\tan\delta = 3.68 \times 10^{-6}$ with $Q_{int} = 2.72 \times 10^5$. By considering the capacitive loss with the participation ratio of the capacitance across MgO among the total capacitance $P_{MgO} \sim 15\%$ and ignoring other losses such as in the junctions as discussed in Ref. 25, the predicted limit of the quality factor of qubit ($Q$) on MgO substrate is estimated to be $Q \sim Q_{int}/P_{MgO} \sim 1.5 \times 10^4$ giving the expected relaxation time $T_1 = Q/(E_{01}/h) \sim 480$ ns for $E_{01}/h$=5 GHz. This limit on $T_1$ set by the dielectric loss in the MgO substrate is indeed comparable to the experimentally observed value of $T_1 \sim 500$ ns.

We can use a similar calculation to estimate the expected quality factor of the qubit grown on a Si substrate. We first consider common sources limiting $T_1$, such as the participation ratio (related to the loss of each material), Purcell effect, and TLS dissipation. Since the Purcell effect limit on $T_1$ can be calculated as $[\kappa \times g^2/(\omega_r - \omega_{01})^2]^{-1} \approx$ 300 μs where $g$ is the coupling strength between resonator and qubit ($g/2\pi \approx 90$ MHz from the experimental data), our results are not limited by the Purcell effect and hence the radiative loss can be ignored. In terms of participation ratio, the overall quality factor of qubit $Q$ is limited by contributions from electric field (capacitive loss: $1/Q_{cap}$), current (inductive loss: $1/Q_{ind}$), and radiative loss ($1/Q_{rad}$) with the relation[36] $1/Q = 1/Q_{cap} + 1/Q_{ind} + 1/Q_{rad}$. In our qubit, the participation ratio of the capacitance across Si ($P_{Si}$) to the total capacitance is about 66.2% ($C_S/C_\Sigma = 52.7$ fF/79.6 fF). If we only consider dielectric (capacitive) loss (i.e. $1/Q \approx 1/Q_{cap}$) and ignore other losses such as loss in the junctions, the quality factor of the qubit is expected to be $Q \approx Q_{cap} = Q_{int}/P_{Si} \sim 4.10 \times 10^5$, which gives the relaxation time $T_1 = Q/(\omega_{01}) \sim 9.88$ μs. This is within a factor of two from the experimentally observed values. Therefore, we ascribe the improvement in $T_1$ (up to the order of tens of μs) to the dominant contribution from the reduced dielectric loss by replacing the MgO substrate with the Si substrate. The enhanced $T_1$ of our qubit due to this contribution also translates into an improvement of $T_2$ compared to that of Ref. 25, since $1/T_2 = 1/2T_1 + 1/T_\varphi$, where $1/T_\varphi$ is the pure dephasing time (estimated to be 63 μs).

**Final remarks.** In order to study the potential of epitaxially grown nitride qubits for superconducting quantum circuits, we have successfully fabricated an all-nitride C-shunt flux qubit epitaxially grown on a Si substrate by utilizing the growth technique of a full-epitaxial NbN/AlN/NbN tri-layer, conventional photolithography with i-line stepper, reactive ion etching, and chemical mechanical polishing (CMP). By employing a Si substrate instead of a conventional MgO substrate in order to reduce the dielectric loss from the substrate, our nitride flux qubit has demonstrated a significant improvement in coherence times, such as $\bar{T}_1 = 16.3$ μs and $\bar{T}_2 = 21.5$ μs, which are more than an order of magnitude longer than those reported in the literature using MgO substrates. We conclude that this improvement in coherence times can be attributed mainly to the reduced dielectric loss when replacing the MgO substrate with the Si substrate. In addition, $T_1$ shows little temporal variation, about 10% deviation from the mean $\bar{T}_1$, indicating that quasiparticles did not strongly influence this device. Although the $\bar{T}_2$ value of 21.5 μs is still far from the record value of 0.3 ms observed in conventional Al-based qubits[15], there remains much room for improvement in the design and fabrication processes of nitride qubits. Considering the advantages of NbN-based superconducting quantum circuits with epitaxial tunnel barriers, our results demonstrate that all-nitride superconducting qubits epitaxially grown on Si substrates have great potential to provide the basis for realizing large-scale superconducting quantum circuits.

## Methods

**Fabrication of all-nitride superconducting qubit.** Samples were fabricated by using epitaxial growth of NbN/AlN/NbN junctions on Si wafers with TiN buffers and planarization by CMP[27,37]. In brief the fabrication flow is as follows. (i) The NbN (200 nm)/AlN (~1.8 nm)/NbN (150 nm) tri-layers are epitaxially grown on a high-resistivity (>20 kΩ·cm) Si substrate with a TiN (50 nm) buffer layer by a dc magnetron sputtering system. (ii) For the device patterning, we used conventional photolithography by an i-line stepper (Canon FPA-3030i5+) and reactive ion etching with $CF_4$ or $CHF_3$ gases. As the first pattern in the device fabrication, the JJ part is formed by etching the upper NbN and AlN layers. (iii) Then the resonator and capacitor patterns are made by etching the lower-NbN and TiN layers. (iv) This is followed by the deposition of a $SiO_2$ film that serves as an isolating layer between the base and wiring layers of the qubit. (v) Then, a planarization process by CMP is carried out to achieve a stable contact hole between the submicron-size JJs and the wiring layer. (vi) After forming contact holes between the base electrode and the wiring layer, the 300-nm-thick NbTiN wiring layer is deposited by dc-magnetron sputtering and patterned. Here the NbTiN wiring layer is chosen because of its smaller superconducting gap ($T_c = 14 - 14.5$ K) compared to that of NbN (~16 K) for the gap engineering so that non-equilibrium quasiparticles are trapped in the wiring layer and do not diffuse into the junction as pointed out in Ref. 25. Since the amorphous $SiO_2$ layer may contain unwanted TLSs, it is removed by etching with BHF just before measurement. From measurements of the current-voltage characteristics of test JJs fabricated on the same wafer as the qubits at 4.2 K, the Josephson critical current density $J_c$, was found to be 40 - 48 A/cm$^2$. Additionally, we confirmed that there is no change in $J_c$ between before and after BHF etching of the SiO$_2$ layer, indicating no damage in the JJs by BHF treatment[37].

**Device parameters.** The fabricated qubit-resonator architecture is depicted in Fig. 1. Our qubit has three JJs, all with circular shapes. Two JJs (i.e., JJ2 and JJ3 in Fig. 1(b)) were designed to have 1.08 μm-diameters (using a mask size of 1.28 μm diameter and expecting a reduction of 0.2 μm after the fabrication process), and the third junction (JJ1 in Fig. 1(b)) was designed to have a 0.70 μm diameter (using a mask size of 0.9 μm diameter) to get a smaller area by a factor $\alpha$ of 0.42. The actual junction diameters after fabrication were 1.07 μm and 0.645 μm, giving the somewhat reduced ratio of $\alpha = 0.36$ as confirmed by scanning electron microscopy images, shown in Fig. 1(c).

The junction and shunt capacitances used in the simulation for the qubit energy spectrum are described in the following. The junction capacitances of NbN/AlN/NbN junctions ($C$) is calculated by the parallel plate capacitor model, i.e. $C = \epsilon_r \epsilon_0 A/t$ where $\epsilon_r$ is the relative permittivity of AlN, $\epsilon_0$ is the permittivity of vacuum, $A$ is junction area, and $t$ is the thickness of insulator (1.8 nm of AlN, which is determined by transmission electron microscopy measurements). The capacitance of the larger junctions in the flux qubit is $C = 31.2$ fF and that of the smaller junction is $\alpha C$. The total junction capacitance of the flux qubit consisting of three junctions is estimated using the relation $C_J = (\alpha + 1/2)C = 26.8$ fF. The small junction is shunted by a shunt capacitance ($C_S$) formed by two rectangular NbN pads. We estimate $C_S \approx 52.8$ fF by considering the same-design C-shunt capacitance fabricated on a sapphire substrate in Ref. 30 (i.e., $C_S = 51$ fF) and the relative permittivity ($\epsilon_r$) 11.5 for sapphire and 11.9 for Si substrates, since the capacitance is proportional to $\epsilon_r$.

Figure 1(d) gives additional information about the thickness profile of the qubit taken from the laser scanning microscope system. Figure 1(e) shows a cross-sectional schematic view of the qubit part indicated by the dashed line in the inset of Fig. 1(b).

For the dispersive readout, the qubit is coupled to a half-wavelength (6.0-mm long) CPW resonator. The center conductor is 10 µm wide, separated from the lateral ground planes by a 6 µm gap, resulting in a wave impedance of the coplanar waveguide $Z = 50$ Ω for optimal impedance matching with conventional microwave components.

**Experimental setup.** The qubit chip was mounted in a sample holder made of gold-plated copper (Fig. 1(a)), which is thermally anchored to the mixing chamber of a dilution refrigerator. For magnetic shielding, the sample holder is covered by a three-layer shield consisting of one aluminum-based superconducting and two µ-metal magnetic shields.

The resonator and qubits are characterized at a base temperature of ~10 mK in a dilution refrigerator. For characterizing the resonator, microwave transmission $S_{21}$ was measured using a vector network analyzer or a heterodyne setup using an IQ mixer and a digitizer. For spectroscopy and coherence measurements of the qubit, we used an additional microwave drive and a commercial analogue-to-digital converter[38], so that the qubit state is read out dispersively via the resonator in a circuit QED architecture.

## Data availability

The data that support the findings of this study are available from the corresponding authors upon reasonable request.

## Acknowledgments

This work was supported by Japan Science and Technology Agency Core Research for Evolutionary Science and Technology (Grant No. JPMJCR1775), JSPS KAKENHI (JP19H05615), JST ERATO (JPMJER1601) and partially by MEXT Quantum Leap Flagship Programs (JPMXS0120319794 and JPMXS0118068682). T.Y. acknowledges the Program for Promoting the Enhancement of Research Universities, Nagoya University.

## Author contributions

All authors contributed extensively to the work presented in this article. S.K., H.T., T.Y., and K.S. designed the experiment and analyzed the data. S.K., W.Q., and H.T. fabricated the samples and characterized the basic junction properties. K.I. performed the microwave measurement. F.Y., T.F., and S.A. supported to data analysis. S.K. wrote the manuscript with feedback from all authors. K.S. and H.T. supervised the whole project.

## Competing interests
The authors declare no competing interests.

## Corresponding authors
Correspondence and requests for materials should be addressed to S.K., H.T. or K.S.

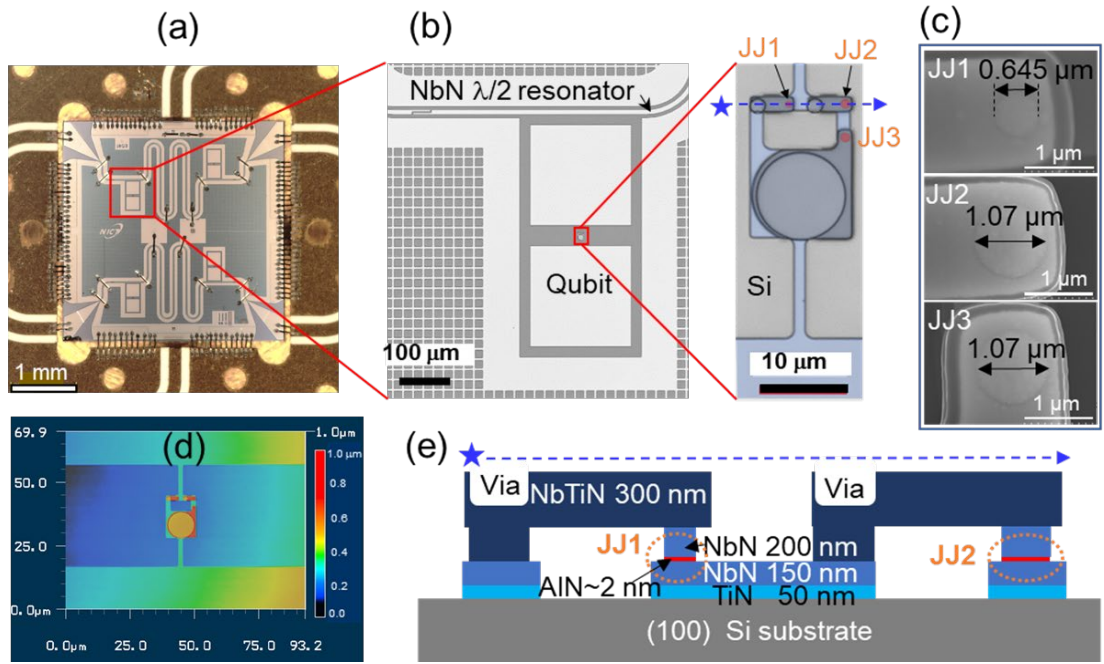

**Fig. 1: All-nitride C-shunt flux qubit consisting of epitaxially grown NbN/AlN/NbN Josephson junctions on Si substrate**. (a) a photograph of the qubit chip mounted into the sample package, (b) Laser scanning microscope image of the capacitively shunted flux qubit coupled to a half-wavelength (λ/2) CPW resonator made of NbN/TiN on a Si substrate. The inset shows a magnified image of a false-colored flux qubit structure with three NbN/AlN/NbN Josephson junctions (marked as JJ1, JJ2, and JJ3). (c) Scanning electron microscopy images corresponding to the three JJs taken after the qubit measurements. (d) The thickness profile of qubit taken from the laser scanning microscope system. The displayed scales are in $\mu$m. (e) Cross-sectional schematics of the part indicated by the blue star and dashed line in (b). The JJ parts are marked by the orange dotted ellipses.

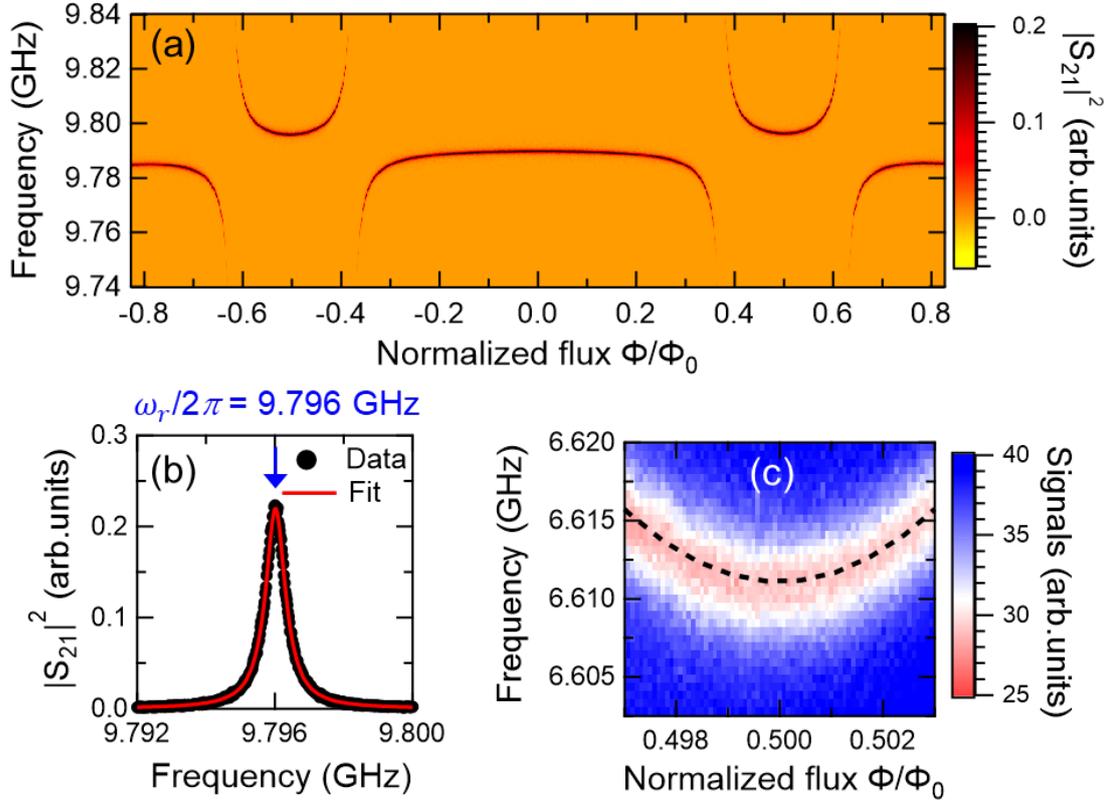

**Fig. 2: Spectroscopy of resonator and qubit.** (a) Spectrum of resonator microwave transmission ($S_{21}$) with varying probe frequency and normalized flux ($\Phi/\Phi_0$). (b) The line profile of the resonator's spectrum at flux bias $\Phi/\Phi_0 = 0.5$ with a Lorentzian fit (solid line). (c) Qubit spectra obtained using dispersive readout. The dashed line is a simulation curve for the qubit transition from the ground state to the excited state ($\omega_{01}/2\pi$). The qubit transition frequency has a minimum value of 6.61 GHz at the flux-insensitive point.

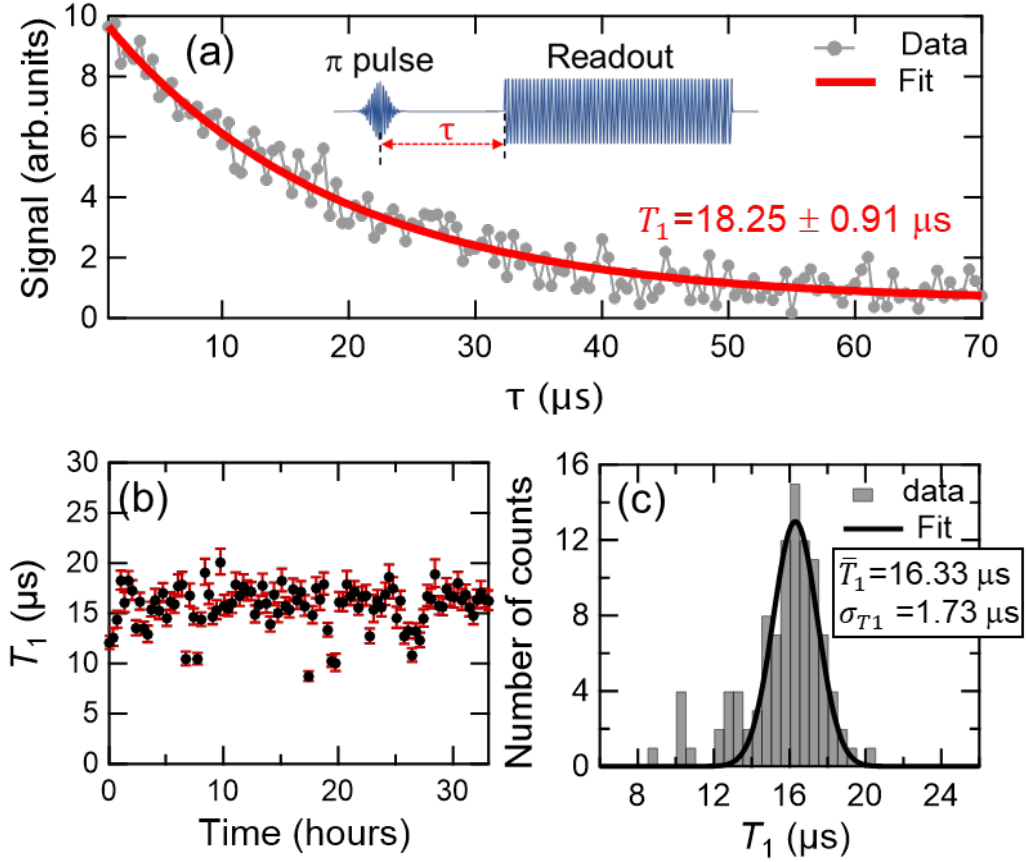

**Fig. 3: Energy relaxation time $T_1$ and its temporal variation.** (a) $T_1$ decay-profile with an exponential fit (solid line) with $T_1 = 18.25 \pm 0.91$ μs. The inset shows the pulse sequence for $T_1$ measurement consisting of a $\pi$ pulse (having a Gaussian envelope) with a 40 ns duration at $\omega_{01}$ and a readout pulse with a 400 ns duration at $\omega_r$. (b) Multiple $T_1$ values obtained from 100 measurements performed over 33 hours, which show the temporal stability of $T_1$. Here the error-bars correspond to the standard deviation calculated in the fitting of each decay profile. (c) Histogram of the $T_1$ values with a Gaussian fit with center value $\bar{T}_1 = 16.3$ μs and standard deviation $\sigma_{T1} = 1.73$ μs (Solid line).

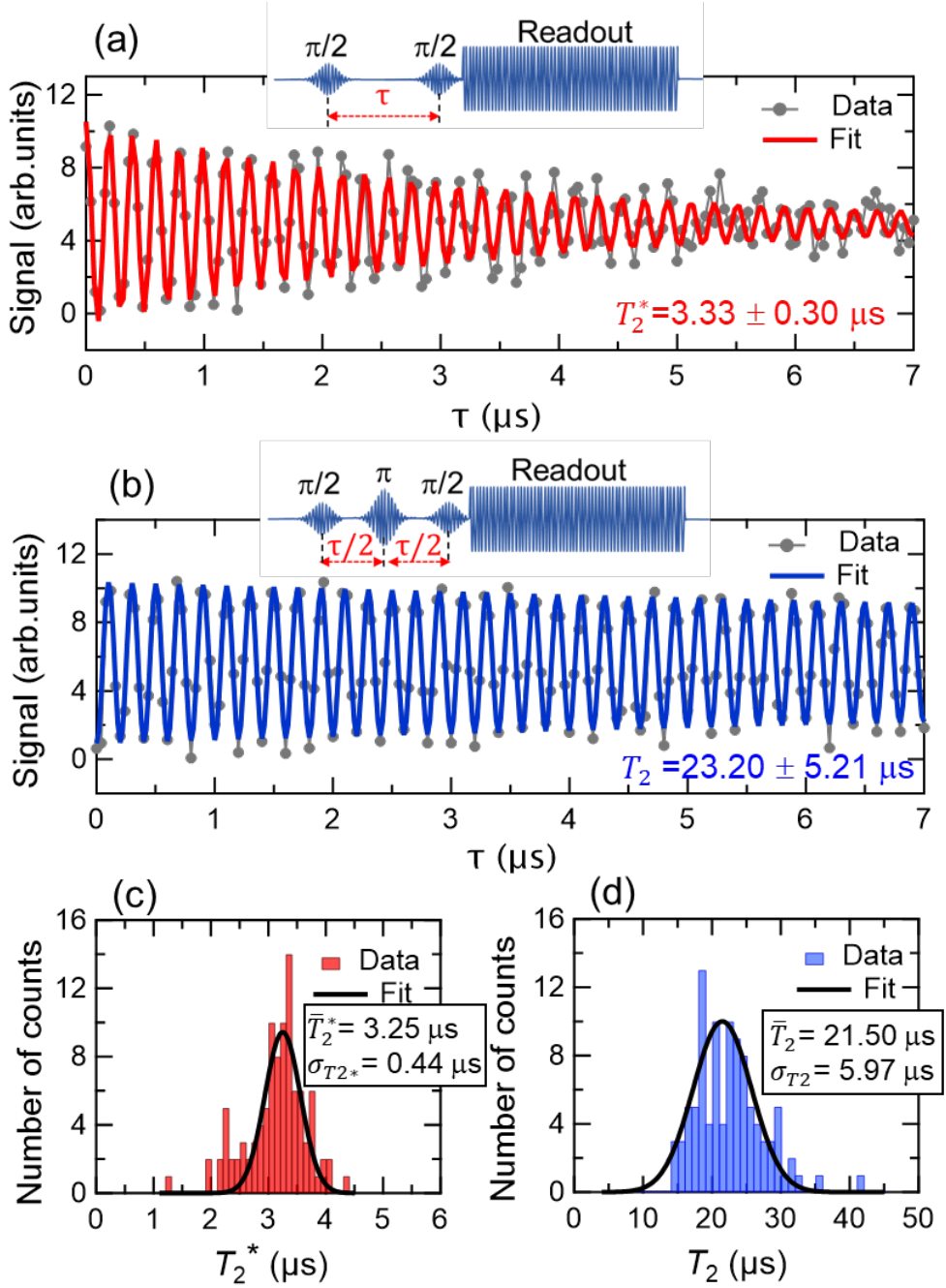

**Fig. 4: Phase relaxation times: $T_2^*$ from Ramsey measurement and $T_2$ from spin-echo experiment.** Time-domain measurements showing (a) Ramsey fringe signal with $T_2^* = 3.33 \pm 0.30$ μs and (b) Hahn echo signal with $T_2 = 23.20 \pm 5.21$ μs. The pulse sequences for the measurements are shown in the insets. The pulse sequences include a $\pi$ pulse with a 40 ns duration, $\pi/2$ pulses with same duration but half the amplitude of the $\pi$ pulses, and readout pulses with 400 ns duration. Here the driving frequency in both measurements is detuned by 5 MHz from the qubit frequency and each coherence time is determined by fitting with an exponentially decaying sinusoidal function. Statistical distributions of (c) the $T_2^*$ values and (d) the $T_2$ values with Gaussian fits (solid line).